 \documentclass[final,5p,times,twocolumn]{elsarticle}

\usepackage{amssymb}

\biboptions{square,numbers,sort&compress}

\journal{Journal of Theoretical Biology}

\begin{document}

\begin{frontmatter}



\title{\Large \textbf{Computational modeling of differences in the
quorum sensing induced luminescence phenotypes of \textit{Vibrio harveyi} and \textit{Vibrio cholerae}}}


\author[VT]{Andrew T Fenley}

\author[Kol]{Suman K Banik}

\author[VT]{Rahul V Kulkarni\corref{cor1}}
\ead{kulkarni@vt.edu}

\cortext[cor1]{Corresponding author}

\address[VT]{Department of Physics, Virginia Polytechnic Institute and State University,
Blacksburg, VA 24061-0435, USA}
\address[Kol]{Department of Chemistry, Bose Institute, 93/1 A P C Road, Kolkata 700 009, India}%

\begin{abstract}
\textit{Vibrio harveyi} and \textit{Vibrio cholerae} have quorum
sensing pathways with similar design and highly homologous components
including multiple small RNAs (sRNAs). However, the associated
luminescence phenotypes of strains with sRNA deletions differ
dramatically: in \textit{V. harveyi}, the sRNAs act additively;
however, in \textit{V. cholerae}, the sRNAs act
redundantly. Furthermore, there are striking differences in the
luminescence phenotypes for different pathway mutants in
\textit{V. harveyi} and \textit{V. cholerae}. However these
differences have not been connected with the observed differences for
the sRNA deletion strains in these bacteria.  In this work, we present
a model for quorum sensing induced luminescence phenotypes focusing on
the interactions of multiple sRNAs with target mRNA. Within our model,
we find that one key parameter -- the fold-change in protein
concentration necessary for luminescence activation -- can control
whether the sRNAs appear to act additively or redundantly.  For
specific parameter choices, we find that differences in this key
parameter can also explain hitherto unconnected luminescence
phenotypes differences for various pathway mutants in
\textit{V. harveyi} and \textit{V. cholerae}.  The model can thus
provide a unifying explanation for observed differences in
luminescence phenotypes and can also be used to make testable
predictions for future experiments.
\end{abstract}


\end{frontmatter}


\newpage

\section{Introduction}
\label{sec:qs-VcVh_intro}

Bacterial survival is critically dependent on regulatory networks
which monitor and respond to environmental fluctuations. An important
example of such a regulatory network is the pathway responsible for
bacterial ``quorum sensing", commonly defined as the regulation of
gene expression in response to cell density \citep{Fuqua1994}.
Quorum sensing bacteria produce, secrete and detect signalling molecules
called autoinducers (AIs) which accumulate in the surroundings as the
cell population increases. Differential expression of certain sets of
genes occurs when the local concentration of AIs exceeds a critical
threshold. Several processes critical to bacterial colonization and
virulence e.g.\ biofilm formation, bioluminescence, and secretion of
virulence factors \citep{Fuqua1996, McFallNgai2000,Miller2001,
Hammer2003, Henke2004a, Waters2005} were shown to be
regulated in this manner, leading to increased interest in
characterizing quorum sensing based regulation in bacteria.

The quorum sensing networks in \textit{Vibrio harveyi} and
\textit{Vibrio cholerae} were recently analyzed in considerable
detail \citep{Ng09}. The basic network components are highly
homologous in the two species, to the extent that the bioluminescence
genes from \textit{V. harveyi}\footnote{The corresponding
bioluminescence genes are absent in \textit{V. cholerae}.} were
used to characterize the regulatory network in \textit{V. cholera}
using luminescence assays \citep{Miller2002,Lenz2004,Lenz2005}.  In
both species, the central regulatory module consists of multiple
quorum regulatory small RNAs (\textit{qrr1-4} in \textit{V. cholerae}
and \textit{qrr1-5} in \textit{V. harveyi}) which control levels of
the master regulator for quorum sensing: LuxR in \textit{V. harveyi}
and HapR in \textit{V. cholerae}. LuxR/HapR levels are maintained
below the threshold for luminescence activation at low cell densities
due to repression by the small RNAs (sRNAs), whereas at high cell
densities quorum sensing leads to a reduction in sRNA production
rates, thereby increasing LuxR/HapR levels above the threshold leading
to luminescence activation. By observing luminescence levels as a
function of cell density for different mutants (corresponding to
different deletions in the quorum sensing pathway components), several
characteristics of pathway structure and function were inferred.

\begin{figure}[h!]
\begin{center}
\begin{tabular}{cc}
\resizebox{42mm}{!}{
\includegraphics{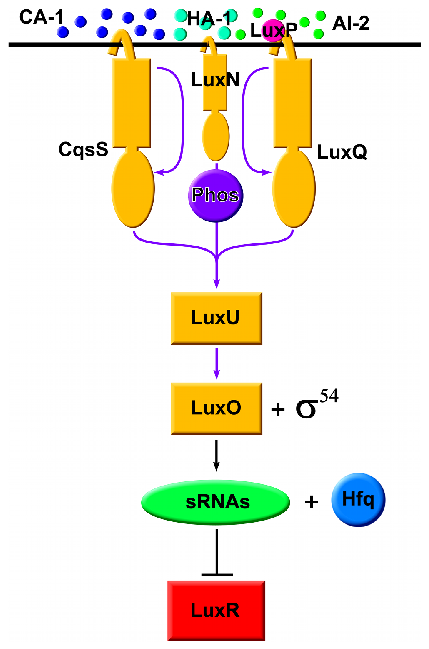}}&
\resizebox{42mm}{!}{
\includegraphics{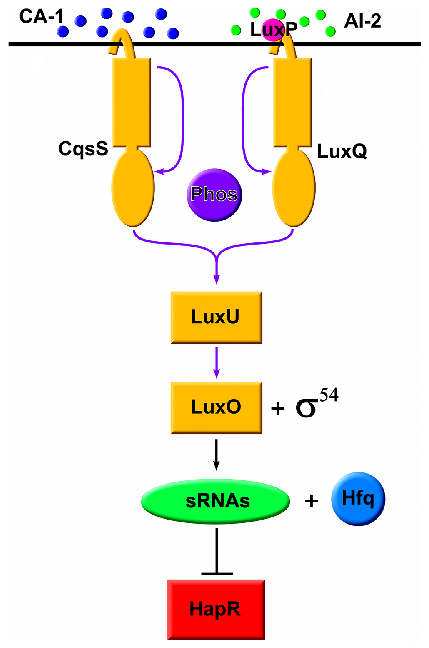}}
\end{tabular}
\end{center}
\caption[The \textit{V. harveyi} and \textit{V. cholerae} quorum sensing gene networks]
    {A schematic of the quorum sensing gene networks.
    (Left) \textit{V. harveyi} and (Right) \textit{V. cholerae}
      employ  multiple AIs whose signals are integrated together
      in order to regulate either LuxR or HapR. \textit{V. harveyi}
      produces and monitors the concentrations of three different AIs
      (HAI-1, CAI-1, and AI-2), while \textit{V. cholerae} produces and
      monitors the concentrations of two different AIs (CAI-1 and AI-2).
      Via very similar phosphorelay networks composed of highly homologous
      components, the bacteria transduce the signal produced by the external
      AI concentrations through the network. In both bacteria, the sensors
      transfer phosphate groups to the protein LuxU when the external
      concentration of AIs is low. LuxU then passes the phosphate groups to
      the protein LuxO which, when phosphorylated, is responsible for the
      production of sRNAs. The flow of phosphate groups slows and then
      reverses when the external concentration of AIs continues to increase,
      thus reducing the production of the sRNAs.}
      \label{fig:Vc_Vh_networks}
\end{figure}

The above studies documented striking differences in luminescence
phenotypes in the two species even though the regulatory components of
the pathways are very similar. The most dramatic differences were seen
in the luminescence phenotypes of the \textit{qrr} sRNA mutants. In
\textit{V. cholerae}, the four \textit{qrr} sRNAs acted redundantly
\citep{Lenz2004} -- all mutants with only one sRNA present
had luminescence phenotypes that were identical to the WT luminescence
phenotype. In contrast, the corresponding sRNAs in \textit{V. harveyi}
acted additively such that different mutants with only one sRNA present
had distinct luminescence phenotypes compared to the WT
phenotype. Thus in \textit{V. harveyi} all the sRNAs must be present
in order to mimic the wild-type luminescence phenotype \citep{Tu2007}.
Apart from these differences in the luminescence phenotypes of the
sRNA mutants, there were also significant differences in the
luminescence phenotypes of strains corresponding to deletions of
upstream pathway elements in the two species. An important challenge
for computational analysis of quorum sensing pathways is to present a
unifying explanation for the various, apparently unrelated,
differences in the luminescence phenotypes for the two species despite
the fact that the pathway elements are very similar (see figure
\ref{fig:Vc_Vh_networks}).

Previous work on modeling quorum sensing globally has
developed a framework for measuring a bacterium's ability to sense its
microenvironment \citep{Pai2009}. By modeling the relative
concentration of autoinducers inside and outside the cell, quorum
sensing bacteria can be characterized based on a `sensing potential'
and its relation to the associated activation properties, e.g.\ the
critical population density \citep{Pai2009}. Modeling has also been
done at the genetic interaction level, in particular the interaction
of sRNAs with target mRNAs. The solutions for the corresponding rate
equations (for a range of parameter values) contained a sharp
transition from a steady state wherein the target mRNA was strongly
repressed to one in which the sRNA was strongly repressed
\citep{Lenz2004,Levine2007,Levine2008,Mehta2008,Mitarai2007,Mitarai2009}.
Since the WT luminescence phenotype also showed a sharp transition as
cell density increased, it was initially suggested that this
transition corresponded to the sharp transition seen in the
sRNA-target rate equations. However, recent experimental results
provide indications that this is not necessarily valid and
correspondingly the picture needs to be revised.

Experiments in {\it V. cholerae} showed that the expression
levels of the virulence regulator AphA \citep{Kovacikova2002} to be
about three-fold lower in WT at low cell densities compared to a
$\Delta hapR$ mutant. This indicates that WT {\it V. cholerae}
maintains HapR at low but significant levels at low cell densities (such
that it can effectively repress AphA to the extent noted) rather
than fully repressing it. Furthermore, experiments in
\textit{V. harveyi} examining regulation of additional targets by LuxR
indicated that LuxR levels change in a graded manner as opposed to a
sharp, ultrasensitive switch \citep{Waters2006}. Thus, there is a need
for computational analysis of sRNA-target regulatory interactions in
the context of the quorum sensing pathway, which is consistent with
these experimental results and which also provides a unifying
explanation for observed luminescence phenotypes.

In what follows, we will present a simplified model for luminescence
regulation during quorum sensing in \textit{V. harveyi} and
\textit{V. cholerae}, which is an extension of work previously done by
the authors of this paper \citep{Banik2009}.  For a given choice of
parameters, the model
accounts for the dramatic differences in the luminescence phenotypes
for the sRNA mutants in the two species based on a single parameter
difference. The analysis also provides a unifying explanation for
currently unrelated differences between the luminescence phenotypes of
different mutants in the quorum sensing pathways and gives rise to
testable predictions for future experiments.
This work thus provides
a framework for systems-level analysis of the quorum sensing pathway
in the \textit{V. harveyi} and \textit{V. cholerae} while complementing
previous models of \textit{V. fischeri}
\citep{Muller2006,Kuttler2008,Muller2008} and suggests future experiments
that can help in further unraveling the function of this critical regulatory
pathway.

\subsection{Overview of experimental results}

We begin with an overview of the two pathways and associated
luminescence phenotypes in the two species. A schematic representation
of the two pathways is shown in figure \ref{fig:Vc_Vh_networks}. The
core elements are the same in both species: a multi-component
phosphorelay involving sensor proteins (which can function as kinases
as well as phosphatases), the phosphotransfer protein LuxU, and the
response regulator protein LuxO. Phosphorylated LuxO is responsible
for the activation of multiple \textit{qrr} sRNAs which in turn
repress the quorum sensing master regulator (LuxR in
\textit{V. harveyi} and HapR in \textit{V. cholerae}).

The pathways do exhibit some differences in the number of autoinducer
synthase/sensor protein pairs and in the number of sRNAs present.
\textit{V. harveyi} has three known autoinducer synthase/sensor protein
pairs whereas \textit{V. cholerae} has only two known autoinducer
synthase/sensor protein pairs. Furthermore, \textit{V. harveyi} has five
\textit{qrr} sRNAs as opposed to four in \textit{V. cholerae} \citep{Ng09}.
However,
our current understanding indicates that these differences are not
significant under the conditions tested. For example, it was shown
that \textit{qrr5} in \textit{V. harveyi} is not quorum sensing
regulated or expressed under normal conditions \citep{Tu2007} and one of the
autoinducer synthase/sensor protein pairs in \textit{V. harveyi} has
minimal effects on quorum sensing based regulation \citep{Henke2004b}.
Thus, both
pathways can effectively be considered as having two autoinducer
synthase/sensor protein pairs and four \textit{qrr} sRNAs.  Furthermore,
the pathway components are highly homologous, e.g. LuxR is greater than
90\% identical to HapR. However, despite these common features and
similarities between components, the luminescence phenotypes show
dramatic differences as detailed below.

\begin{figure}[h!]
\begin{center}
\resizebox{75mm}{!}{
\includegraphics{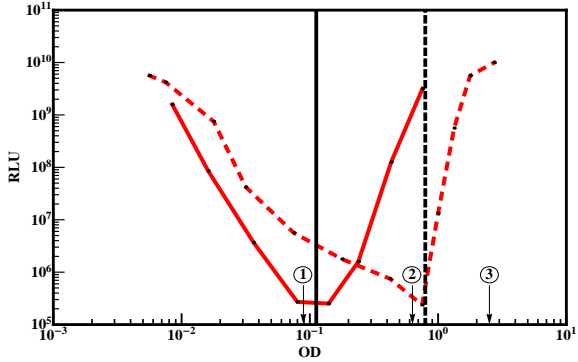}}
 \caption[Wild-Type luminescence curves for \textit{V. harveyi}
 and \textit{V. cholerae}]
 {Experimental Wild-Type luminescence curves for \textit{V. harveyi}
 and \textit{V. cholerae}. The solid, red curve represents the change in
 luminescence relative to optical density (OD) for \textit{V. harveyi}. There
 is a smooth transition in luminescence near OD $10^{-1}$ as the
 distribution of cells switch from ``off" to ``on" \citep{Tu2007}. The dashed,
 red curve represents the change in luminescence relative to OD for
 \textit{V. cholerae}. There is a sharp transition in luminescence near OD
 $10^0$ as the distribution of cells switch from ``off" to ``on"
 \citep{Lenz2004}. The vertical solid and dashed lines represent possible OD
 concentrations that correspond to the beginning of the cells in the population
 reaching a LuxR/HapR concentration necessary for luminescence for
 \textit{V. harveyi} and \textit{V. cholerae} respectively.
 Regions indicated
 by (1), (2), and (3) reflect the relative protein distributions labeled similarly
 and shown in figure \ref{fig:threshold_WTQS}.
}
\label{fig:WT_RLU}
\end{center}
\end{figure}

While the most dramatic differences in luminescence
phenotypes occur in the qrr and \emph{luxU} mutants, distinct
qualitative differences exist even in the WT phenotypes, see figure
\ref{fig:WT_RLU}. These qualitative differences serve as motivation
for our modeling and help in discerning a key difference between the
two genetic networks that can potentially explain the various
differences in luminescence phenotypes.
The luminescence curves for WT strains of \textit{V. harveyi} and
\textit{V. cholerae} (based on experimental data from \citep{Tu2007}
and \citep{Lenz2004}) are shown in figure \ref{fig:WT_RLU}. In both
cases, the luminescence per cell begin at a high value since the
initial state corresponded to a dilution of the high cell density
culture which was maximally bright. As the colony density increases,
the luminescence level drops until a critical cell density is reached,
after this critical point there is a subsequent rise in luminescence
back to the initial level. While the luminescence curves of WT
\textit{V. harveyi} and \textit{V. cholerae} look similar, there are
important differences between the two curves. Wild-type
\textit{V. harveyi} showed an almost symmetric parabola centered
around OD$_{600} \sim$ 0.1 \citep{Tu2007}; however, wild-type
\textit{V. cholerae} showed a continued decline in relative light unit
(RLU) output until the colony reached an OD$_{600}$ $\sim$ 1.0. The
luminescence levels then increased by several orders of magnitude over
a timescale during which cell density changed by a small factor
($\approx 4$ fold) \citep{Lenz2004}.

The luminescence phenotypes of strains corresponding to deletions of
various pathway elements also depicted important differences between the
two species. As mentioned in the Introduction, luminescence curves of
{\it qrr} sRNA mutants in the two species suggested that the sRNAs
functioned additively in \textit{V. harveyi} \citep{Tu2007} but were
redundant in \textit{V. cholerae} \citep{Lenz2004}. Another striking
difference was seen in the \textit{luxU} mutant which was always bright
regardless of cell density in \textit{V. harveyi} whereas the \textit{luxU}
mutant showed a density-dependent luminescence phenotype in \textit{V. cholerae}. 
Furthermore, while deletion of the sensor kinases (e.g.\ for the
\textit{cqsS},\textit{luxQ} mutant) changed the luminescence phenotype
with respect to WT for \textit{V. harveyi}, the corresponding WT and deletion
mutant luminescence curves were almost identical for
\textit{V. cholerae} \citep{Lenz2005}. These observations based
on experimental luminescence curves lead to some important questions
which need to be addressed:

\noindent
1) How can we understand changes in RLU (Relative Light Unit)/cell over
several orders of magnitude corresponding to small changes in cell density?
\noindent
2) How are the phenotypes dramatically different despite the basic
components/circuitry being the same?
\noindent
3) Is there a unifying explanation for the seemingly unrelated differences
in luminescence phenotypes for different mutant strains?


\section{Methods}
\label{sec:qs-VcVh_methods}
\subsection{Modeling framework}

In order to address the issues raised above, we will first discuss the
modeling framework and key assumptions of our model. They are
schematically illustrated in this section and more quantitatively
developed in following sections.

We assume that the measured luminescence levels per cell are
proportional to the rate of transcription of the luminescence genes.
Since these genes are activated by the quorum sensing master
regulators, their transcription rate is a function of cellular
concentrations of LuxR/HapR. We assume that this function has a sharp
threshold; as a simplification we represent it by a step function such
that cells with LuxR/HapR concentrations below the threshold produce
no light whereas cells with LuxR/HapR concentrations above the
threshold produce maximal luminescence.

Since the experimentally
measured quantity is the population average of the luminescence
output/cell, we need to consider the steady state distribution of
LuxR/HapR levels across all cells. Recent work showed that the
steady state protein distribution for proteins can be characterized as
a Gamma distribution \citep{Friedman2006}. Accordingly, we represent
the LuxR/HapR distribution by a Gamma distribution with a given
variance and whose mean value is determined by solving the rate
equations of our model (see next section).

\begin{figure}[h!]
\begin{center}
\resizebox{75mm}{!}{
\includegraphics{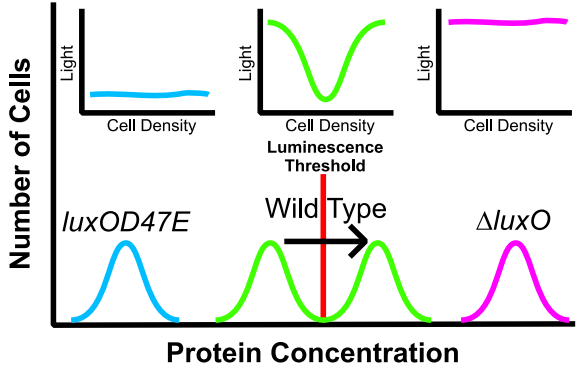}}
\caption[Luminescence activation due to a sharp threshold]
{An illustration depicting luminescence activation as
LuxR/HapR concentrations cross a sharp threshold for activation. More
(less) than the threshold, luminescence is (not) activated.
The different distributions depicted in this illustration are
examples of LuxR/HapR concentration distributions and the
corresponding luminescence profile for a few examples. (left - cyan) A
protein distribution that remains below the threshold regardless of
cell density. (middle - green) A protein distribution that transitions
across the threshold and is a function of cell density. (right -
magenta) A protein distribution that is entirely past the threshold
regardless of cell density.}
\label{fig:threshold_cartoon}
\end{center}
\end{figure}

With the assumptions mentioned above, we can make significant
inferences about quorum sensing networks based on the luminescence
data. The change in RLU/cell over several orders of magnitude
corresponds to the steady state distribution for LuxR/HapR crossing
the luminescence activation threshold (see figure
\ref{fig:threshold_cartoon}). Thus the mean concentration of LuxR/HapR
must change by the minimal amount indicated in the figure during the
transition from the `dark' phenotype to the maximally luminescent
phenotype. The WT luminescence curves indicate that this change occurs
gradually in \textit {V. harveyi} (positions (1) and (2) in figure
\ref{fig:WT_RLU}) as compared to \textit {V. cholerae} (positions (2) and (3)
in figure \ref{fig:WT_RLU}). Since the change in mean HapR levels in
\textit{V. cholera} (at OD $\sim 1.0$) occurs without a corresponding
significant change in cell density, it is unlikely to be driven solely
by quorum sensing. Instead we infer, based on the luminescence
phenotype, that there is a sharp rise in HapR levels around OD $\sim
1.0$ in \textit{V.cholerae}.
One potential cause for this rise is a further
reduction in the available regulatory sRNAs allowing for more available
\textit{hapR} transcripts. A possible source of the sRNA reduction is that as
the cells move into
stationary phase from growth phase, there is a decrease in the production of
the Hfq chaperone \citep{Ishihama1999}. A decrease in Hfq corresponds to
a decrease in the concentration of sRNA-Hfq complexes which are necessary
to regulate the target mRNA. Recent experiments in
\textit{V. cholerae} have indeed found evidence for a sharp rise in
HapR levels at OD $\sim 1.0$ \citep{Svenningsen2008}.

In contrast, the transition in the
WT luminescence phenotype for \textit{V. harveyi} occurs at lower OD
values and is more gradual suggesting that it is driven by the
quorum sensing pathway. This observation
leads to the suggestion that the crucial difference between the two
species lies in the location of the threshold for luminescence
activation: in \textit{V. harveyi}, quorum sensing based regulation
suffices for moving the steady state LuxR distribution across the
threshold, whereas in \textit{V. cholerae} this requires an additional
jump in HapR levels at OD $\sim 1.0$.

\subsection{A minimal model for luminescence activation}

We focus on quorum sensing pathway elements corresponding to the
production of sRNAs, the transcription of the target mRNA
(\textit{luxR} or \textit{hapR}), and the interaction between the
sRNAs and target mRNA. We start with a model containing only one sRNA
species and neglect autoregulation of the target protein. Then we add
the contributions of multiple sRNAs and autoregulation to the model.

The basic equations for a simplified model of sRNA-target interaction
have been introduced and analyzed in previous work
\citep{Lenz2004,Levine2007,Levine2008,Mitarai2009} and are given below
(equations (\ref{eq:simple_x}) and (\ref{eq:simple_y})).  Consider
first the case of a single sRNA species regulating one target mRNA
species. If $[x]$ denotes the concentration of the sRNA and $[y]$ the
concentration of the target mRNA, the corresponding equations are:
\begin{eqnarray}
            \label{eq:simple_x}
            \frac{d[x]}{dt}&=&k_x - \gamma [x][y] - \mu_x [x], \\
            \label{eq:simple_y}
            \frac{d[y]}{dt}&=& k_y - \gamma [x][y] - \mu_y [y],
        \end{eqnarray}

\noindent where the $k$'s are the production rates of each species, the
$\mu$'s are the degradation rates of each species, and $\gamma$ is
an effective parameter for mutual degradation of sRNA and target mRNA.

To generalize the above equations (\ref{eq:simple_x}) and
(\ref{eq:simple_y}) while taking care of the effective parameter constraints
(see Appendix A), we include the effects
of 1) multiple sRNAs regulating {\it luxR}/{\it hapR} and 2) autoregulation of
LuxR/HapR \citep{Chatterjee1996,Lin2005}. The corresponding equations are,
 \begin{eqnarray}
\label{eq:multiple_auto_x}
\frac{d[x_i]}{dt} &=& k_{x_i} - \gamma_i [x_i][y] - \mu_{x_i} [x_i], \\
\label{eq:multiple_auto_y}
\frac{d[y]}{dt} &=& \frac{k_y}{1+({[y}/{[y_D]})} - \sum_i \left(\gamma_i [x_i][y]\right) - \mu_y [y] ,
\end{eqnarray}

\noindent The constant ${[y_D]}$, represents the threshold
concentration for binding of the target protein to its own mRNA.
When the target protein is
bound to the promoter region, transcription of the target gene is
effectively blocked.

Bioinformatic analysis \citep{Lenz2004} indicates that the 32 bp
region in the {\it qrr} sRNAs which is involved in regulation of {\it
hapR/luxR} is absolutely conserved for all the sRNAs. Thus, we make
the assumption that all the sRNAs have the same affinity for the
target mRNA, i.e.\ we set $\gamma_{i} = \gamma $. We further assume that
the degradation rates of all sRNAs are the same ($\mu_{i} = \mu$).
However, the model does consider differences in the sRNAs production
rates ($k_{x_{i}}$) as demonstrated by experiment \citep{Tu2008}.

At steady state, the mean protein concentration is the mean mRNA
concentration scaled by a constant -- the ratio of the protein translation
rate to the protein degradation rate. Therefore, we use the scaled
mRNA concentration in place of the protein concentration (see Appendix).

To make the connection to luminescence curves, we have to consider the
distribution of protein levels across cell populations.  Recent work
by Friedman \textit{et al}.\ showed the distribution of the protein
concentration per cell for the colony can be represented by a
Gamma distribution
\citep{Friedman2006}. Furthermore, recent flow cytometry work showed the
distributions of fluorescence per cell from a \textit{luxR-gfp} fusion had a
nearly constant variance for a variety of conditions related to the
concentration of AIs \citep{Waters2006}. Similarly, luminescence output from
Vibrio harveyi was shown to be heterogenous across cell populations
\citep{Anetzberger2009}. Therefore, we model the
protein distribution as a Gamma distribution with a fixed variance.
The mean of the distribution is obtained from the equations above for a
given choice of parameters. Using this framework, we show in the following
section how a single parameter difference can account for the vastly different
luminescence phenotypes of {\it V. harveyi} and of {\it V. cholerae}.

\section{Results and Discussion}

In this section we show how the minimal model discussed above with only
one essential difference (the threshold for luminescence activation)
between the \textit{V. harveyi} and \textit{V. cholerae} pathways can
explain the observed differences in luminescence phenotypes as well as
lead to testable predictions.

We note that bacterial colonies are observed to change their
luminescence production by many orders of magnitude in a relatively
short amount of time, see figure \ref{fig:WT_RLU}. However, the
changes in the level of the master regulator proteins and sRNAs are
not nearly as dramatic \citep{Waters2006,Tu2007}. We interpret this as
indicating that a significant fraction of all the cells in the colony
reach the conditions necessary for luminescence activation upon a
small change in the master regulator protein levels. We
model this as corresponding to a significant fraction of the master
regulator distribution moving across sharp threshold values of
concentrations necessary to activate luminescence (see figure
\ref{fig:threshold_cartoon}).

\subsubsection{The protein distributions for WT strains}

\begin{figure}[ht!]
\begin{center}
\resizebox{75mm}{!}{
\includegraphics{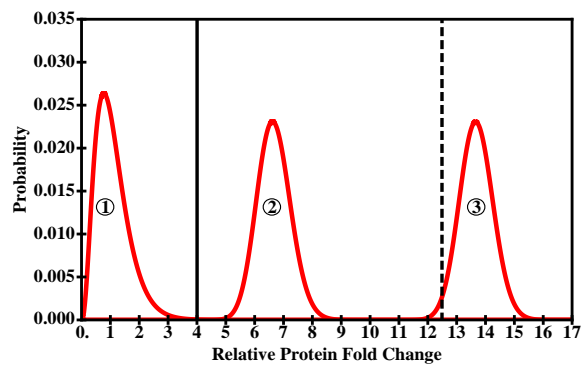}}
\caption[Distributions of the protein concentration across a WT
bacterial colony at different density limits] {The distributions of
the protein concentration across a WT bacterial colony from the model
for the: (1) low-cell density limit, (2) high-cell density limit, and
(3) entering stationary-phase limit. The x-axis depicts the fold
change difference relative to the mean protein concentration value for
a WT colony at the low-cell density limit. The solid, vertical bar
between distributions (1) and (2) and the dashed, vertical bar
vertical between distributions (2) and (3) represent the threshold
values for luminescence for \textit{V. harveyi} and
\textit{V. cholerae}, respectively.}
\label{fig:threshold_WTQS}
\end{center}
\end{figure}

Using the model equations with parameter values guided by experiment
(see Appendix),
we plot the the distribution of the protein concentration for a WT colony
(representing either \textit{V. harveyi} or \textit{V. cholerae}) at the
low-cell density limit, high-cell density limit, and entering stationary
phase labeled as positions (1), (2), and (3) respectively, see figure
\ref{fig:threshold_WTQS}.
Since the protein distributions for WT and all the mutants in either
\textit{V. harveyi} or \textit{V. cholerae} are not available, we plot the
distributions with respect to fold changes relative to the mean protein
concentration for a WT colony at the low-cell density limit.\footnote{At the
time of this work, the experimental protocol for measuring the exact
protein concentration per cell \textit{in vivo} was not available and has now
only recently been published \citep{Teng2010}.} Specifically,
the first two distributions in figure \ref{fig:threshold_WTQS} are
representative of the maximal relative change in protein concentration
in going from low-cell density to high-cell density based on changes due
to quorum sensing alone. The third distribution in figure
\ref{fig:threshold_WTQS} is the resulting distribution after the
final reduction in sRNA production leading to a rise in HapR due to
entering stationary phase.

The distributions at positions (1) and (2) in figure
\ref{fig:threshold_WTQS} represent the maximally dark and maximally
bright WT \textit{V. harveyi} colonies, respectively. Similarly, the
distributions at positions (1-2) and (3) in figure
\ref{fig:threshold_WTQS} represent the maximally dark and maximally
bright WT \textit{V. cholerae} colonies, respectively. Ideally none of
the bacteria in the dark colony should be ``on" and none of the
bacteria in the bright colony should be ``off", therefore we set the
threshold of light activation for \textit{V. harveyi} at a fold change
directly in between the two distributions -- depicted as the solid,
vertical line in figure \ref{fig:threshold_WTQS}. Since experiments
have shown that the activation of \textit{V. cholerae} to occur at a
larger cell density than \textit{V. harveyi}, we propose the threshold
of light activation for \textit{V. cholerae} to be at a larger fold
change -- depicted as the dashed, vertical line in figure
\ref{fig:threshold_WTQS}.  As indicated in the figure, this
corresponds to luminescence activation occurring in
\textit{V. harveyi} using quorum sensing alone, whereas for
\textit{V. cholerae} luminescence activation requires both transition
to the high-cell density limit for the quorum sensing pathway and
additional changes in HapR levels associated with entry into
stationary phase.  In what follows, we will discuss how assuming
\textit{V. cholerae} has a different threshold of light activation
than \textit{V. harveyi} can consistently explain the differences in
the sRNAs and $luxU$ mutant phenotypes.

\subsubsection{Additivity vs redundancy}

\begin{figure}[ht!]
\begin{center}
\resizebox{75mm}{!}{
\includegraphics{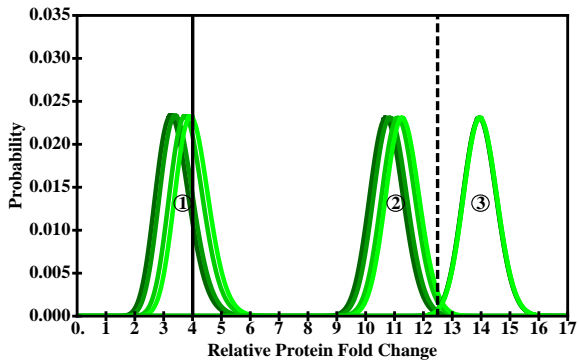}}
\caption[Distributions of the protein concentration across a mutant
colony containing only one active qrr sRNA from our model at different
density limits]
{The distributions of the protein concentration across a
mutant colony containing only one active qrr sRNA from our model at the:
(1) low-cell density and (2) high-cell density limits. The x-axis depicts the fold
change difference relative to the mean protein concentration value for
a WT colony at the low-cell density limit. The solid, vertical bar between
distributions (1) and (2) and the dashed, vertical bar vertical between
distributions (2) and (3) represent the threshold values for luminescence
for  \textit{V. harveyi} and \textit{V. cholerae} respectively.}
\label{fig:threshold_qrrsQS}
\end{center}
\end{figure}

We account for each of the four active qrr sRNAs having a different
production rate and set the rates with the following hierarchy:
qrr4 $>$ qrr2 $>$ qrr3 $>$ qrr1, which is consistent with experimental
results in \textit{V. harveyi} \citep{Tu2007}. Figure \ref{fig:threshold_qrrsQS}
shows the distributions of the protein concentrations for mutant
colonies containing only one of the four active qrr sRNAs for both
\textit{V. harveyi} and \textit{V. cholerae} -- each sRNA mutant is
represented as a different shade of green in
figure \ref{fig:threshold_qrrsQS}.

In the low-cell density limit, position (1) in figure
\ref{fig:threshold_qrrsQS}, the distributions all have regions
extending past the threshold for luminescence in
\textit{V. harveyi}. This is a representation of the qrr ``additivity"
response seen in \textit{V. harveyi} as all qrrs are needed to prevent
any appreciable region of the protein distribution from extending past
the threshold in the low-cell density limit \citep{Tu2007}. For the
sRNA mutants, the regions of the distributions in the low-cell density
limit that extend past the threshold represent the amount of bacteria
in the colony that are ``on" regardless of cell density.

The story is different from the perspective of
\textit{V. cholerae}. In the high-cell density limit, position (2) in
figure \ref{fig:threshold_qrrsQS}, the distributions are all below the
threshold for luminescence in \textit{V. cholerae}, which corresponds
to complete light repression and mimics the WT \textit{V. cholerae}
response \citep{Lenz2004}.  We suggest that once the final reduction in
sRNA production occurs, e.g.\ entering stationary phase, all the
distributions cross the threshold for luminescence, position (3) in
figure \ref{fig:threshold_qrrsQS}. The resulting phenotype looks to be
the same as the WT \textit{V. cholerae} phenotype with the conclusion
that the sRNAs act ``redundantly". However, the prediction from our
model is that the sRNAs behave the same in both \textit{V. harveyi}
and \textit{V. cholerae}, but the associated thresholds for
luminescence are different.

\subsubsection{The luxU mutant}

The \textit{luxU} mutant is another example of a difference in
luminescence phenotypes between \textit{V. harveyi} and
\textit{V. cholerae}.  The protein LuxU is responsible for coupling
the autoinducer input signal to the rest of the quorum sensing
network, see figure \ref{fig:Vc_Vh_networks}. If LuxU is removed from
the pathway, the total sRNA transcription rate would drop to minimal
levels, and the system would no longer respond to changes in cell
density. Therefore if the quorum sensing pathway is the only factor
controlling the luminescence phenotypes, removal of the the $luxU$
gene should result in a bright, density independent phenotype.  For
\textit{V. harveyi}, this is indeed the case -- the $luxU$ mutant is
bright regardless of cell density.

The story, as before with the sRNAs, is different with
\textit{V. cholerae}.  In \textit{V. cholerae}, the $luxU$ mutant
shows a density dependent luminescence phenotype, but the shape of the
luminescence curve is different from the canonical quorum sensing
luminescence curves \citep{Miller2002}. In the low-cell density limit,
there is a detectable level of light production that is larger than WT
value or any of the sRNA mutants values but much less than the maximal
level of light production. This low level of luminescence remains
stable for a significant portion of the exponential phase, and then
sharply increases to the maximum level of luminescence -- a feature
present in most \textit{V. cholerae} luminescence curves.

\begin{figure}[h!]
\begin{center}
\resizebox{75mm}{!}{
\includegraphics{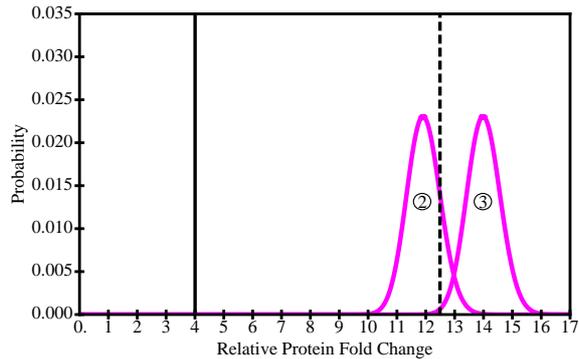}}
\caption[Distributions of the protein concentration across a mutant
colony not containing \textit{luxU} at different density limits]
{The distributions of the protein concentration across a
mutant colony where \textit{luxU} has been removed from the system from
our model at the:
(2) high-cell density limits and (3) high-cell density limit entering
stationary phase. The x-axis depicts the fold change difference
relative to the mean protein concentration value for a WT colony at
the low-cell density limit. The solid, vertical bar between
distributions (1) and (2) and the dashed, vertical bar vertical between
distributions (2) and (3) represent the threshold values for luminescence
for  \textit{V. harveyi} and \textit{V. cholerae} respectively.}
\label{fig:threshold_luxU}
\end{center}
\end{figure}

Our model reproduces this observed $luxU$ mutant behavior in
\textit{V. harveyi} and \textit{V. cholerae}. In figure
\ref{fig:threshold_luxU}, there are only two distributions: one for
the high-cell density limit (position (2)) and one for the high-cell
density limit entering stationary phase (position (3)).  From the
perspective of LuxR/HapR regulation, the removal of $luxU$ effectively
decouples the quorum sensing pathway from the outside inputs.
Therefore, the system effectively starts at the high-cell density
limit, and the associated protein distribution is always past the
\textit{V. harveyi} luminescence threshold. This results in a fully
bright, density independent phenotype, see figure
\ref{fig:threshold_luxU}. However, the distribution associated with
the high-cell density is only partially across the
\textit{V. cholerae} luminescence threshold resulting in a small
concentration of the cells being ``on" and a majority being
``off". The \textit{V. cholerae} colony will remain in this state
until it enters stationary phase where the protein distribution
completely crosses the \textit{V. cholerae} luminescence threshold,
see figure \ref{fig:threshold_luxU}.

Finally, the luminescent behavior of the $cqsS$ and $luxQ$ double mutant
in \textit{V. cholerae} is also consistent with the model. Essentially, this
double mutant shows a WT response even though the both autoinducer
sensors are removed \citep{Lenz2005}. In our model, this would correspond
to the system starting in the WT high-cell density limit, position (2) in figure
\ref{fig:threshold_WTQS}, which is below the threshold for luminescence in
\textit{V. cholerae}. Therefore, the observed phenotype should be nearly
identical to WT.


By just having two thresholds separating three distinct regions of
protein regulation, our model is able to consistently link the sRNAs
acting additively in \textit{V. harveyi} \citep{Tu2007}, the sRNAs
acting redundantly in \textit{V. cholerae} \citep{Lenz2004}, and the
density dependent phenotype in \textit{V. cholerae} for the $luxU$
mutant \citep{Miller2002}. With the relative positions of the
thresholds and protein distributions now in place, we now discuss the
predictions that come from our model.

\subsection{Predictions}

Current experimental techniques can produce a variety of different
mutant strains of \textit{V. harveyi} and of
\textit{V. cholerae}. Depending on the genes and sRNAs being removed
from the strain, the experimental techniques generate even up to triple
knock-out mutants (and possibly more if required). Since our simplified
model, with the given choice of parameters, reproduces features of the
observed luminescence phenotypes, it is of interest to examine the
model predictions for luminescence phenotypes of
different gene and sRNA mutant strains that should be experimentally
feasible to test.


The model distinguishes the varying behaviors of \textit{V. harveyi}
and of \textit{V. cholerae} as a difference in the threshold protein
concentration of the master regulatory gene, and the concentration of
the master regulatory gene at any position is determined by the
associated production rate of the sRNAs. Therefore, there is an
effective total sRNA production rate that coincides with the
distribution of the master regulatory protein being centered at a
given threshold value. We refer to this critical value of total sRNA
production as $k_c$.

Since we assume the threshold values are different for
\textit{V. harveyi} and \textit{V. cholerae}, their associated
critical value of total sRNA production, $k_c$, is different. Each
mutant has an associated total sRNA production rate at low-cell
density and high-cell density limits. A hierarchy of sRNA production
rates for different mutants and colony cell densities relative to
$k_c$ explains currently seen phenotypes, and we will use this
hierarchy as a basis for predicting new phenotypes.
\begin{eqnarray}
\label{eq:hierarchy_harveyi}
WT_{l} > sRNA_{l} > k_c >WT_{h} > sRNA_{h} > \Delta U >  \Delta O, \\
\label{eq:hierarchy_cholerae}
WT_{l} > sRNA_{l} > WT_{h} > sRNA_{h} > \Delta U > k_c > \Delta O .
\end{eqnarray}

\noindent Equations (\ref{eq:hierarchy_harveyi}) and
(\ref{eq:hierarchy_cholerae}) represent the hierarchies for
\textit{V. harveyi} and \textit{V. cholerae}, respectively. Those
rates greater than $k_c$ correspond to ``dark" phenotypes, and those
rates less than $k_c$ correspond to ``bright" phenotypes.  In
equations (\ref{eq:hierarchy_harveyi}) and
(\ref{eq:hierarchy_cholerae}), $WT_{l}$ and $WT_{h}$ represent the
total sRNA production rate for wild-type bacteria in the low-cell
density and high-cell density limits before the transition to
stationary phase. $sRNA_{l}$ and $sRNA_{h}$ are the sRNA production
rates for any mutant with at least one active sRNA removed from the
system in the low-cell density and high-cell density limits before the
transition to stationary phase. Finally, $\Delta U$ and $\Delta O$ in
equations (\ref{eq:hierarchy_harveyi}) and
(\ref{eq:hierarchy_cholerae}) are the sRNA production rates for the
mutants where LuxU and LuxO has been deleted, respectively. Now that
the hierarchy is established, we discuss below the resulting
predictions.

One way to explore the different quorum sensing responses of the
network is to add an external concentration of autoinducers to a
low-cell density colony, also know as ``cross-feeding". The additional
autoinducers will ``trick" a colony into behaving as if it is in the
high-cell density limit which causes a transition in the total sRNA
production rate. Since the production rates specifically dependent on
cell density, $WT_{l}$, $sRNA_{l}$, $WT_{h}$, and $sRNA_{h}$ are
separated by $k_c$ in equation (\ref{eq:hierarchy_harveyi}), the model
predicts a low-cell density colony of wild-type or any sRNA mutant
\textit{V. harveyi} will start to luminesce when extra autoinducers
are added to the colony.

However, for \textit{V. cholerae}, the production rates specifically
dependent on cell density are all greater than $k_c$ in equation
(\ref{eq:hierarchy_cholerae}). Even the production rate of the $luxU$
mutant is greater than $k_c$. Therefore, the model predicts that a
low-cell density colony of wild-type or any sRNA mutant
\textit{V. cholerae} will remain dark when extra autoinducers are
added to the colony. Also, the model predicts this outcome for any
mutant \textit{V. cholerae} corresponding to a total sRNA production
rate greater than $k_c$, including the $luxU$ mutant and the $cqsS$ and
$luxQ$ double mutant.

The model also predicts cases where the sRNAs act ``additively" in the
$luxU$ mutant \textit{V. cholerae}. Since the total sRNA production
rate associated with the $luxU$ mutant (\ref{eq:hierarchy_cholerae})
is adjacent to $k_c$, reducing the total sRNA production rate to a
value less than $k_c$ will result in light production. This could be
achieved by combining $luxU$ with sRNA mutants. Therefore, the
different sRNA triple mutants in combination with the $luxU$ mutant
for \textit{V. cholerae} should show the associated HapR concentration
changing in a graded manner. Thus our model predicts that, in a
$luxU$ mutant background, the different triple sRNA mutants will
appear to behave additively with regards to the luminescence --
a phenotypic response similar to sRNA mutants in WT \textit{V. harveyi}.

\subsection{Discussion}

In this study, we have shown how a simple set of equations, with appropriate
choice of parameters, can effectively mimic the quorum sensing luminescence
phenotypes of
\textit{V. harveyi} and \textit{V. cholerae}. While the components of the
quorum sensing regulatory network in each of the bacteria are biologically
similar in both homology and function, there are striking differences in
luminescence phenotypes for the same mutant,
e.g.\ $luxU$. Even the sRNAs, which are virtually identical
in their sequence specificity to the target gene, act additively versus
redundantly for \textit{V. harveyi} and \textit{V. cholerae}.

We account for the striking differences by suggesting that the
threshold concentration of the master protein needed for the bacteria
to start luminesce activation is larger in \textit{V. cholerae} than
\textit{V. harveyi}.  The larger threshold concentration correspondingly
implies the need for a mechanism that increases the levels
of the master regulator in addition to the increases due to
quorum sensing. The increase in master regulator levels can be effectively
modeled as a sharp drop in sRNA productions rates and one possible
source of this reduction can arise from the transition from
exponential growth phase to stationary phase.

We considered solutions of the model equations for specific parameter choices
motivated by experiments and analyzed the effect different mutants have on
the sRNAs' production
rates. In \textit{V. harveyi}, the removal of either LuxO or LuxU
causes a sufficient reduction in the sRNAs' production rate to result
in the bacterial colony achieving maximal luminescence at any cell densities.
Only the removal of
LuxO from \textit{V. cholerae} results in a similar response. Removing
LuxU does not drop the sRNAs' production rates enough for the bacteria
to luminesce at any cell density. The extra reduction in the sRNAs'
production rates from the transition to stationary-phase is required
for the $luxU$ mutant of \textit{V. cholerae} to luminesce.

We note that differences in the rate parameters between
the two species could also account for some of the observed differences
in phenotypes. However, quantifying the multitude of reaction rates in both
organisms is a challenging task experimentally. Also, even if significant
differences in
the rates between the two species are found, it is not clear if they
will account for the observed dramatic differences in phenotypes.
Instead of considering a multitude of parameter differences as the
explanation for the observed phenotypes, our model suggests that
changes in a single parameter can provide a unifying explanation for all the
observed differences. Furthermore our model leads to testable
predictions which can easily be validated experimentally (e.g.\ changes
in luminescence upon crossfeeding).

We further note that an alternative explanation for the
differences in the luminescence phenotypes (in particular for the
$luxU$ mutant phenotypes) is based on the observation that the
VarA/S-CsrA pathway interacts with the the quorum sensing pathway in
\textit{V. cholerae} \citep{Lenz2005}. While the corresponding
interaction has not been studied in \textit{V. harveyi}, it is likely
that VarA/S-CsrA pathway interacts with the quorum sensing pathway in
a similar fashion in \textit{V. harveyi}. If, however, it turns out
that the interaction between the two pathways is absent in {\it
V. harveyi}, then this observation could account for some of the
differences seen. In this case, the fold-change required for
luminescence activation would be higher in \textit{V. cholerae} due to
repression by CsrA taking place in \textit{V. cholerae} but not in
\textit{V. harveyi}. However, even with the deletion of $csrA$ in
\textit{V. cholerae}, the $luxU$ mutant still shows a
density-dependent phenotype in \textit{V. cholerae} \citep{Lenz2005} in
contrast to the observed phenotype in \textit{ V. harveyi}.  Thus
there is a clear difference between the $luxU$ mutant phenotypes
between the two species which cannot be ascribed to potential
differences in the interactions between the quorum sensing pathways
and the VarA/S-CsrA pathway.

Within our model, the relationship between the threshold concentration
and the total of all the sRNAs' production rates leads to experimental
predictions. The first prediction is the inability to prematurely
initiate luminescence in a low-cell density colony of
\textit{V. cholerae} through the addition of a large concentration of
autoinducers.  Thus cross-feeding based activation of luminescence
should work in \textit{V. harveyi} but not in \textit{V. cholerae} We
also predict that a $luxU$ mutant of \textit{V. cholerae} combined
with sRNAs mutants will result in a phenotype where the sRNAs act
additively.

In summary, we have presented a simplified model for quorum sensing
induced luminescence phenotypes in \textit{V. harveyi} and
\textit{V. cholerae}. Our analysis suggests that a single parameter
difference in our model effectively reproduces many features of observed
luminescence curves which were hitherto unconnected. Thus large sequence-based
differences are, in principle, not required to explain the dramatic
differences between the luminescence phenotypes in these two species.
Our model also makes testable predictions for observable
luminescence phenotypes (specifically in \textit{V. cholerae}) which,
if validated, should shed new light on luminescence regulation by
quorum sensing.


\section*{Acknowledgements}

RVK acknowledges support from the U.S National Science Foundation
(NSF) through grant PHY-0957430. ATF acknowledges support from NSF
IGERT grant DGE-0504196.  The authors also acknowledge support
from ICTAS (Virginia Tech) and the Virginia Tech ASPIRES award.


\section{Appendix}
\label{sec:appendix}

\subsection{Single sRNA model}

Here we provide additional details for the single sRNA model.
For convenience, we introduce the following dimensionless parameters;
$\tilde{x} = (\mu_x / k_x) [x]$, $\tilde{y} =  (\mu_y / k_y) [y] $,
$\alpha = (\gamma k_y)/(\mu_x \mu_y)$, and
$\beta = (\gamma k_x)/(\mu_x \mu_y)$ in equations
(\ref{eq:simple_x}) and (\ref{eq:simple_y})
so that the corresponding equations at steady state become:
 \begin{eqnarray}
            \label{eq:dimensionless_simple_x}
            0 &=& 1 - \alpha \tilde{x}\tilde{y} - \tilde{x}, \\
            \label{eq:dimensionless_simple_y}
            0 &=& 1 - \beta \tilde{x}\tilde{y} - \tilde{y}.
        \end{eqnarray}

\noindent These equations can be readily solved to determine how
steady state sRNA-mRNA levels change as system parameters are
varied. In the limit $\alpha, \beta \gg 1$, the solutions show a sharp
transition as the ratio $\alpha / \beta$ changes from $(\alpha /
\beta) < 1$ to $(\alpha / \beta) > 1$. This parameter regime lets the
system respond in an \emph{ultrasensitive} manner as discussed in
previous works \citep{Lenz2004,Mitarai2009}. During quorum sensing, the
production rate of the sRNA ($k_{x}$) decreases and hence the
parameter $\beta$ is lowered as bacteria make the transition from low-cell
density to high-cell density.  Correspondingly the system evolution
traces out a trajectory in $(\alpha,\beta)$ phase space.  For $\alpha,
\beta \gg 1$ the target mRNA levels show a sharp change as the line $
\alpha = \beta$ is crossed; thus it seems natural to identify the
sharp transition observed in the luminescence profile with the sharp
transition in target mRNA levels as $\beta$ is lowered. However, as
argued in the previous sections, this identification is unlikely to be
valid based on the following observations: 1) the quorum sensing
response in {\it V. harveyi} is observed to be graded rather than
all-or-none \citep{Waters2006}.  2) Recent experiments have shown that
HapR represses aphA at low cell density \citep{Kovacikova2002}, thus
target mRNA levels are significant even at low cell densities. 3)
northern blots show little difference in the amount of sRNA in
\textit{V. cholerae} when the target mRNA ({\it hapR}) is deleted
\citep{Lenz2004}.

Observation 3.) from above suggests that the sRNA and mRNA interactions
occur in a parameter regime where the sRNA is never fully suppressed.
To adhere to this constraint, we look at the limit $\alpha \ll 1$ and $\beta >
1$ which effectively holds the sRNA concentration constant regardless of
the target mRNA concentration. In this limit, the system no longer has an
ultrasensitive response, but instead responds in a controlled manner. As sRNA
production rates double, the mRNA concentrations are about halved,
thus allowing for a graded response.
Applying the limits $\alpha \ll
1$ and $\beta > 1$ to equations (\ref{eq:dimensionless_simple_x}) and
(\ref{eq:dimensionless_simple_y}) in the steady state explicitly shows
the controlled response:
\begin{eqnarray}
\label{eq:dimensionless_ss_x}
\tilde{x} &\approx& 1,\\
\label{eq:dimensionless_ss_y}
\tilde{y} &\approx& 1/(1+\beta) .
\end{eqnarray}

We note that recent research \citep{Svenningsen2008} has
provided evidence for dosage compensation in \textit{V. cholerae} due to
regulation of sRNA production by HapR. While the inclusion of this effect
will imply that the restrictions on the parameter $\alpha$ noted above are
not required, the observation that the response of the regulated target is
graded in a controlled manner remains the same.



\subsection{Multiple sRNA with autoregulation model}

Here we provide additional details for the multiple sRNA with
autoregulation model. In \textit{V. harveyi}, there are a total of five
sRNAs; however, only four are actively controlling the concentration
of \textit{luxR} mRNA. Likewise, \textit{V. cholerae} contains four active
sRNAs. Including multiple sRNAs has generated the new constants:
$k_{x_i}$, $\gamma_i$, and $\mu_{x_i}$. However, we make
the assumptions that each sRNA has equal affinity to the target mRNA
and all the sRNAs have the same degradation rate in both
bacteria. These assumptions return $\gamma_i$ to $\gamma$ and
$\mu_{x_i}$ to $\mu_x$. To model autoregulation, we introduce the
dimensionless parameter $\tilde{y_D}$ as the threshold concentration for
effective autoregulation of the target gene. In dimensionless units, the
system should be tuned in such a way that $\tilde{y_D}$ is not larger than
the maximum value obtainable by $\tilde{y}$ which is 1.

Using similar dimensionless parameters as the single sRNA model, we
replace $\beta$ with $\beta_i = (\gamma k_{x_i})/(\mu_x \mu_y)$,
and introduce the dimensionless parameter $\epsilon=\mu_y / \mu_x$,
which is only necessary in the time dependent solutions of the model.
Equations (\ref{eq:multiple_auto_x}) and (\ref{eq:multiple_auto_y}) are
therefore rewritten as the following set of dimensionless equations
\begin{eqnarray}
\label{eq:dimensionless_multiple_auto_x}
\epsilon \frac{d\tilde{x}_i}{d\tilde{t}} &=& 1 - \alpha \tilde{x}_i \tilde{y} - \tilde{x}_i ,\\
\label{eq:dimensionless_multiple_auto_y}
\frac{d\tilde{y}}{d\tilde{t}} &=& \frac{1}{1+(\tilde{y}/\tilde{y}_D)} -
\sum_i\beta_i \tilde{x}_i\tilde{y} - \tilde{y} .
\end{eqnarray}


The addition of multiple sRNAs to the model does not change the
production rate of the target mRNA; therefore, we still consider the system to
be in the parameter space where $\alpha \ll 1$. At steady state,
$\tilde{x}_i \approx 1$ and the summation in equation
(\ref{eq:dimensionless_multiple_auto_y}) reduces to $\sum_i\beta_i
\tilde{y}$. Since $\tilde{y}$ is independent of the summation, the sum
is only of $\beta_i$, which results in just a constant representing
all the contributions of the sRNAs, $\sum_i\beta_i \rightarrow
\beta_{total}$. The effects of multiple sRNAs are all integrated into
the constant $\beta_{total}$, and their removal via mutations to the
wild-type bacteria is equivalent to reducing the maximum and minimum
value of $\beta_{total}$ as the bacteria moves from low cell density
to high cell density respectively. The steady state concentration of
$\tilde{y}$ therefore becomes:
\begin{equation}
\tilde{y}\left(1 + \frac{\tilde{y}}{\tilde{y}_D}\right) = \frac{1}{1 + \beta_{total}} .
\label{eq:dimensionless_ss_poly_auto}
\end{equation}

%

%

\noindent
The effect of the autoregulation is best seen via different limiting
cases of ratio $\tilde{y}/\tilde{y}_D$ in equation
(\ref{eq:dimensionless_ss_poly_auto}). When $\tilde{y}_D \ll 1$, then
$\tilde{y}/\tilde{y}_D \gg 1$ resulting in $\tilde{y} \approx
\sqrt{\tilde{y}_D/(1 + \beta_{total})} \approx 0$. This corresponds
to the case where autoregulation is maximally on which prevents the system
from sustaining any appreciable amount of protein. When $\tilde{y}_D \gg 1$, then $\tilde{y}/\tilde{y}_D \ll 1$ resulting in $\tilde{y} \approx
1/(1 + \beta_{total})$ which is similar in form to equation
(\ref{eq:dimensionless_ss_y}) where autoregulation is absent from
the system. Since the amount of $\tilde{y}$ is constrained to a value
between 0 and 1, and $\tilde{y}_D$ is the effective concentration needed of
the target protein before autoregulation occurs, we set  $\tilde{y}_D = 0.75$,
which corresponds to the production rate dropping by close to half as seen by experiment \citep{Chatterjee1996}.

\subsection{Parameter space analysis}

Here we discuss the various parameter values and their associated
experimental motivation used in the preceding models. Fluorescence
experiments involving the expression of \textit{V. harveyi}'s qrr2 in the
low-cell density and high-cell density limits provide a possible measure
for estimating the sRNA fold change between the two cell density limits
\citep{Waters2006}.  The same type of fluorescence experiment shows
the translational rate of \textit{luxR} \citep{Waters2006} when LuxR
autoregulation is removed. A direct determination of relative fold
differences using Real-Time Quantitative PCR for qrr1, qrr2, qrr3, qrr4,
qrr5, and \textit{luxR} with autoregulation intact has also been done
\citep{Tu2007}. In the case without autoregulation, \textit{luxR}
translational levels change $\sim$10 fold and qrr2 expression levels
also change $\sim$10 fold.


\begin{table}
    \label{table:mutant_rates}
    \begin{center}
    \begin{tabular}{lccc}
    \hline
        & LCD (1) & HCD (2) & Stationary (3) \\
    \hline
    WT      & $\beta_i(f_{LCD}) = 20.4$ & $\beta_i(f_{HCD}) = 2.04$ & $\delta\beta_i(f_{HCD}) = 0.051$\\
    \emph{qrr1}  & $\beta_1(f_{LCD}) = 4.5$ & $\beta_1(f_{HCD}) = 0.45$ & $\delta\beta_1(f_{HCD}) = 0.0125$\\
    \emph{qrr2}  & $\beta_2(f_{LCD}) = 5.4$ & $\beta_2(f_{HCD}) = 0.54$ & $\delta\beta_2(f_{HCD}) = 0.0135$\\
    \emph{qrr3}  & $\beta_3(f_{LCD}) = 4.8$ & $\beta_3(f_{HCD}) = 0.48$ & $\delta\beta_3(f_{HCD}) = 0.0120$\\
    \emph{qrr4}  & $\beta_4(f_{LCD}) = 5.7$ & $\beta_4(f_{HCD}) = 0.57$ & $\delta\beta_4(f_{HCD}) = 0.0143$\\
    $\Delta$\emph{luxU} & $\beta_i(f_O) = 0.3264$ & $\beta_i(f_O) = 0.3264$ & $\delta\beta_i(f_O) = 0.00816$\\
    $\Delta$\emph{luxO} & $\beta_i(0) = 0.0$ & $\beta_i(0) = 0.0$ & $\delta\beta_i(0) = 0.0$ \\
    \hline
    \end{tabular}
    \end{center}
    \caption[The different $\beta_i(f)$ values used in the model]{A table of
    the different $\beta_i(f)$ values for WT, $\Delta$\emph{luxU},
    $\Delta$\emph{luxO}, and the \emph{qrr} mutants.}
\end{table}

With regards to the experimentally shown constraints, we let $\beta_i$
change 10 fold between the low-cell density and high-cell density
limits. Since we are in the limit where $\beta_i$ is always greater
than 1, we chose $\beta_i \approx 20$ for the low-cell
density resulting in $\beta_i \approx 2$ for the high-cell density
limit. We set $\alpha = 0.1$ to satisfy the previously discussed
constraint: $\alpha \ll 1$. The values chosen for $\alpha$ and
$\beta_i$ minimally satisfy the limits set on parameter space; and
yet, the system behaves in a manner consistent with
experiment. Smaller values of $\alpha$ and/or larger values of
$\beta_i$ are also consistent with experiment showing robustness of
the system in this parameter regime.

To incorporate the effect of the system entering stationary phase, we
introduced the parameter $\delta$ such that $k_{x_i}
\rightarrow \delta k_{x_i}$. we set $\delta$ to the fixed value 0.025 --
the maximal value necessary to have a clear enough distinction between
the distributions at position (2) and (3) for the $luxU$ mutant, where
position (3) represents the colony entering stationary phase (see
figure \ref{fig:threshold_luxU}).

$\alpha$, $\beta_i$, and $\delta$ are the only parameters necessary
to determine the (normalized) mean values of LuxR/HapR. Furthermore,
$\alpha$ and $\delta$ remain a fixed value throughout our analysis, 0.1
and 0.025 respectively. $\beta_i$, which is a function of the sRNA
production rates, only changes in value between the low-cell density limit
and the high-cell density limit. The effects of the different mutants are
also embedded into $\beta_i$ as they represent variations to the sRNA
production rates relative to the WT.


The critical factor in determining the decomposition of $\beta_i$ is the
fraction of LuxO ($f$) that is capable of promoting the production of sRNA.
Therefore, $\beta_i$ is a function of $f$, $\beta_i(f)$. To better understand
the contributions of LuxU and LuxO to $\beta_i(f)$, we specify the different
values $\beta_i(f)$ can achieve depending on cell density and genotype.
First, quantitative real-time PCR experiments show a basal rate sRNA
production that is independent of the presence of LuxO which we label:
$\beta_i(0)$ \citep{Tu2007}. Next there is the rate, $\beta_i(f_O)$, that
depends on the presence of LuxO which is evident in the $\Delta luxU$
mutant showing a wild-type like luminescence phenotype in \textit{V.
cholerae} \citep{Miller2002}. Then there are the rates associated with
phosphorylating LuxO, the dominant factor in sRNA production, in the low
cell density limit ($\beta_i(f_{LCD})$) and in the high cell density limit
($\beta_i(f_{HCD})$). The different values $\beta_i(f)$ for WT,
$\Delta$\emph{luxU}, $\Delta$\emph{luxO}, and the \emph{qrr} mutants
 are listed in table~1.

\bibliographystyle{elsart-harv}
\bibliography{QS_Paper_JTB_Bib}







\end{document}